\newcommand{\PRE}[1]{}      
\documentclass[aps,twocolumn,amsmath,preprintnumbers,amsfonts,amssymb]{revtex4}
\usepackage{bm}
\usepackage{epsfig}
\usepackage{graphicx}

\newcommand{\be}{\begin{equation}}
\newcommand{\ee}{\end{equation}}
\newcommand{\bea}{\begin{eqnarray}}
\newcommand{\eea}{\end{eqnarray}}
\newcommand{\nn}{\nonumber}
\newcommand{\ba}{\begin{array}} 
\newcommand{\ea}{\end{array}}

\newcommand{\larray}[2][lllll]{\begin{array}{#1} #2\\ \end{array}}

\newcommand{\lsim}{
\mathrel{\hbox{\rlap{\hbox{\lower4pt\hbox{$\sim$}}}\hbox{$<$}}}}
\newcommand{\gsim}{
\mathrel{\hbox{\rlap{\hbox{\lower4pt\hbox{$\sim$}}}\hbox{$>$}}}}

\newcommand{\md}{\text{~mod~}}

\begin{document}

\preprint{UFIFT-HEP-08-03}

%-----------------------------------
% Title
%-----------------------------------
\title{
\PRE{\vspace*{1.5in}}
Lightest $\boldsymbol{U}$-parity Particle (LUP) dark matter
\PRE{\vspace*{0.3in}} }
\author{Hye-Sung Lee}
\affiliation{
Institute for Fundamental Theory, University of Florida, Gainesville, FL 32611, USA}
\PRE{\vspace*{.1in}}
\date{February, 2008}

\begin{abstract}
\PRE{\vspace*{.1in}} \noindent
We suggest a $U(1)'$ gauge symmetry as an alternative to the usual $R$-parity of supersymmetric standard models, showing that it can also work as a common source of stabilities of proton and dark matter in addition to other attractive features.
The residual discrete symmetries of a single $U(1)'$ can provide stabilities to both the MSSM sector (proton) and the hidden sector (new dark matter candidate, LUP).
The LUP can expand the viability of many models such as $R$-parity violating models and gauge mediation models regarding dark matter issue.
\end{abstract}

\pacs{11.30.Fs, 12.60.Jv, 95.35.+d}
%11.30.Fs   Global symmetries (e.g., baryon number, lepton number)
%12.60.Jv   Supersymmetric models
%95.35.+d   Dark matter

\maketitle

%=======================================================================
% MAIN TEXT
%=======================================================================

%--------------------------------------------------------------------------------
\section{Introduction}
%--------------------------------------------------------------------------------
It is now an established fact that matters in our Universe is composed of visible world as well as dark world.
Large scale structures of both worlds depend on stabilities of building blocks such as proton and dark matter.

TeV scale supersymmetry (SUSY) is a well-motivated new physics scenario which resolves the gauge hierarchy problem of the standard model (SM).
Mere supersymmetrization of the SM allows lepton number ($\cal L$) and baryon number ($\cal B$) violations at renormalizable level, and there is no guaranteed stabilities for proton and dark matter candidate.
Further, the $\mu$-problem associated with two Higgs doublets arises \cite{Kim:1983dt}.
While a complete solution to theses issues may exist only at higher scale physics such as grand unified theory, it would be worth seeking if a stand-alone TeV scale physics model can be constructed describing the world without apparent problems.
We take this bottom up approach and discuss the SUSY companion symmetry manifest at TeV scale that can resolve the problems of SUSY models.

$R$-parity is a strong candidate for this companion symmetry since it addresses stabilities of building blocks of both worlds with one discrete symmetry, and the $R$-parity conserving minimal supersymmetric standard model (MSSM) has been the most popular TeV scale SUSY model.
However, $R$-parity lacks some features to be a fulfilling TeV scale SUSY companion symmetry.
While the $R$-parity provides absolute stability to the lightest superparticle (LSP) dark matter candidate, it still allows too fast proton decay with dimension five operators (such as $QQQL$ and $U^cU^cD^cE^c$) \cite{Weinberg:1981wj}.
Further, it forbids both $\cal L$ violating and $\cal B$ violating terms completely at renormalizable level, which would be unnecessary if the dark matter is not the LSP.

TeV scale Abelian gauge symmetry $U(1)'$ \cite{Langacker:2008yv} may be a phenomenologically more attractive companion symmetry for the TeV scale SUSY model.
The $\mu$-problem can be solved very naturally \cite{UMSSM}, and it can further prevent dimension five proton decay operators.
Though one can adopt both the $R$-parity and the $U(1)'$ together, it would be more economical and desirable if one symmetry can address all aforementioned issues.
While the $R$-parity itself may be embedded in the TeV scale $U(1)'$, we will consider the $R$-parity violating case to contrast the $U(1)'$ with the $R$-parity and to fully exploit the experimentally allowed possibilities.

$R$-parity violating terms are allowed while ensuring the proton stability among the MSSM fields due to the automatic {\em LV-BV separation} found in the model, which prevents coexistence of the $\cal L$ violating terms and $\cal B$ violating terms \cite{Lee:2007fw}.
When a $U(1)'$ gauge symmetry is introduced, it may bring two other things: residual discrete symmetries and exotic fields for anomaly cancellation.
The exotics in general may regenerate fast proton decay \cite{Weinberg:1981wj}, but with help of the residual discrete symmetries of the model identified in Ref. \cite{Lee:2007qx}, the proton stability can be ensured even with TeV scale exotics.
However, decay of the LSP in the absence of the $R$-parity may be a serious shortcoming of the model since the dark world stability is not guaranteed.

In Ref. \cite{Hur:2007ur}, a residual $Z_2$ discrete symmetry ($U$-parity) of the $U(1)'$ was proposed as a discrete symmetry among the fields which are singlet under the $G_\text{SM} = SU(3)_C \times SU(2)_L \times U(1)_Y$.
The lightest $U$-parity particle (LUP) is stable, and is a new dark matter candidate.
An independent $R$-parity was assumed in Ref. \cite{Hur:2007ur} for the proton stability at renormalizable level on top of the $U$-parity.
It was numerically illustrated that this multiple (LSP and LUP) dark matters scenario can explain both relic density and direct detection constraints easily\footnote{For another example of multiple dark matters analysis, see Ref. \cite{Cao:2007fy}.}.

In this letter, we consider the $U(1)'$ as a common source of the discrete symmetries in both observable sector ($G_\text{SM}$ charged) and hidden sector ($G_\text{SM}$ uncharged), which can ensure the stabilities for the proton and the hidden sector dark matter candidate (LUP).
Due to the gauge origin, these remnant discrete symmetries are not violated by the Plank scale physics \cite{Krauss:1988zc}.
Thus, the $U(1)'$ which interacts with both the observable and hidden sectors can serve as a good TeV scale SUSY companion symmetry that can provide a unified solution to the $\mu$-problem, proton stability, and dark matter stability, without the $R$-parity.

%--------------------------------------------------------------------------------
\section{Discrete symmetries of the observable sector}
%--------------------------------------------------------------------------------
We define the observable sector (or the MSSM sector) as the $G_\text{SM}$ charged fields and three right-handed neutrinos.
The right-handed neutrino ($N^c$) does not carry the $G_\text{SM}$ charges, but it can form a single particle coupled with left-handed neutrino (e.g. $\frac{\left<S\right>}{M} H_u L N^c$ can be a light Dirac neutrino \cite{Langacker:1998ut}).
The other $G_\text{SM}$ singlet fields will be defined as hidden sector, which can still have the $U(1)'$ charges.
In this letter, we will not consider any other gauge symmetries in the hidden sector.

The renormalizable superpotential of the MSSM sector is given by
\bea
&&W_\text{MSSM}  = \mu H_u H_d + y^D_{jk} H_d Q_j D^c_k + y^U_{jk} H_u Q_j U^c_k \nn \\
&&~~~~ + y^E_{jk} H_d L_j E^c_k + y^N_{jk} H_u L_j N_k^c + [ \lambda_{ijk} L_i L_j E^c_k \nn \\
&&~~~~ + \lambda'_{ijk} L_i Q_j D^c_k + \mu'_i H_u L_i + \lambda''_{ijk} U^c_iD^c_jD^c_k ] \label{eq:WMSSM}
\eea
where the bracketed part does not respect $R$-parity.
We do not specify any exotic fields that might exist with $G_\text{SM}$ charges, since they are model-dependent.
A Higgs singlet $S$ breaks the $U(1)'$ spontaneously, and its vacuum expectation value (vev) can serve as effective coefficients replacing the original coefficients of the superpotential (see Sec. \ref{sec:beyond}).

The possible discrete gauge symmetries for the MSSM fields from an extra $U(1)$ symmetry were investigated in Refs. \cite{Ibanez:1991pr,Dreiner:2005rd,Luhn:2007gq}.
A residual discrete gauge symmetry $Z_N$ emerges if the discrete charges ($q[F_i]$) and the $U(1)'$ charges ($z[F_i]$) satisfy the following relation:
\be
z[S] = N, \quad z[F_i] = q[F_i] + n_i N
\ee
for each field $F_i$.
The Higgs singlet $S$ is supposed to have $q[S] = 0$ to keep the discrete symmetry unbroken after the $U(1)'$ symmetry is spontaneously broken by $S$ (e.g. both $S F_1 F_2$ and $\left<S\right> F_1 F_2$ should be singlet under the discrete symmetry).

The most general $Z_N$ discrete symmetry of the MSSM can be written as (using the basis $B_N \equiv R_N L_N$ instead of usual $R_N$)
\be
Z_N^{obs}:~~ g_N^{obs} = B_N^b L_N^\ell \label{eq:BLbasis}
\ee
with family-universal cyclic symmetries ($\Phi_i \to e^{2\pi i \frac{q_i}{N}} \Phi_i$)
\be
B_N = e^{2\pi i \frac{q_B}{N}}, \quad L_N = e^{2\pi i \frac{q_L}{N}} .
\ee
The discrete charges ($q_B$, $q_L$) of each generator are listed in Table \ref{tab:discrete}, and the total discrete charge of $Z_N^{obs}$ is $q = b q_B + \ell q_L \md N$.
As Table \ref{tab:discrete} shows, the discrete charges of $B_N$ and $L_N$ are closely related to baryon number and lepton number, respectively, and a general discrete charge can be written as $q = -b {\cal B} - \ell {\cal L} + b (y/3) \md N$ with a conserved quantity of $-(b {\cal B} + \ell {\cal L}) \md N$. % and the hypercharge.
For example, for $R_N = B_N L_N^{-1}$,
\be
q_R = q_B - q_L = -({\cal B - L}) + (y/3) \md N .
\ee

%==================== TABLE ====================================================================
\begin{table}[tb]
\begin{center}
\begin{tabular}{|l||r|r|r|r|r|r|r|r|r|l|}
\hline
               & $Q$   & $U^c$ & $D^c$ & $L$   & $E^c$ & $N^c$ & $H_u$ & $H_d$ & $X$            & ~~meaning of $q$ \\
\hline
\hline
$L_N$          & $0$   & $0$   & $0$   & $-1$  & $1$   & $1$   & $0$   & $0$   & $0$              & $-{\cal L}$ \\
$B_N$          & $0$   & $-1$  & $1$   & $-1$  & $2$   & $0$   & $1$   & $-1$  & $0$              & $-{\cal B} + y/3$ \\ 
$R_N$          & $0$   & $-1$  & $1$   & $0$   & $1$   & $-1$  & $1$   & $-1$  & $0$              & $-({\cal B-L}) + y/3$ \\ 
$U_2$          & $~~0$ & $~~0$ & $~~0$ & $~~0$ & $~~0$ & $~~0$ & $~~0$ & $~~0$ & $-1$  & $-{\cal U}$ \\
\hline
\hline
$~y$           & $1$   & $-4$  & $2$   & $-3$  & $6$   & $0$   & $3$   & $-3$  & $0$            & \\
\hline
\hline
$L_{3N}^3$     & $0$   & $0$   & $0$   & $-3$  & $3$   & $3$   & $0$   & $0$   & $0$              & $-3{\cal L}$ \\ 
$B_{3N}^3 - y$ & $-1$  & $1$   & $1$   & $0$   & $0$   & $0$   & $0$   & $0$   & $0$              & $-3{\cal B}$ \\
$R_{3N}^3 - y$ & $-1$  & $1$   & $1$   & $3$   & $-3$  & $-3$  & $0$   & $0$   & $0$              & $-3({\cal B-L})$ \\
\hline
\end{tabular}
\end{center}
\caption{
Discrete charges ($q$) of the $L_N$, $B_N$, $R_N$ and their equivalent $Z_{3N}$ via scaling and hypercharge ($y$) shift as well as the discrete charge of $U_2$.
\label{tab:discrete}}
\end{table}
%=============================================================================================

%--------------------------------------------------------------------------------
\section{Discrete symmetries extended to the hidden sector}
%--------------------------------------------------------------------------------
Now, what is the phenomenologically favored discrete symmetry?
The matter parity ($R_2 = B_2 L_2^{-1}$), which is equivalent to the $R$-parity has been the most popular choice.
It provides absolute stability to the LSP, but proton is not sufficiently stable, and the $\mu$-problem requires another mechanism.
The baryon triality ($B_3$) provides absolute stability to the lightest baryon (proton) regardless of possible heavy exotic fields\footnote{Proton decay (which requires $\Delta {\cal B} = 1$) is not allowed by the selection rule of $B_3$ ($\Delta {\cal B} = 0 \md 3$) \cite{Castano:1994ec}.}, but does not prevent too fast LSP decay\footnote{Of course, one can consider combinations of discrete symmetries such as $R_2 \times B_3$ \cite{Dreiner:2005rd}.}.
There are also other discrete symmetries such as $L_3$ and $B_3 L_3$ that can ensure sufficient proton stability with additional conditions \cite{Lee:2007qx}.

However, the dark matter does not have to belong to the MSSM sector, and it may be a hidden sector particle that are charged under the $U(1)'$ symmetry.
In this letter, we consider the simplest case given by
\be
W_\text{hidden} = \frac{\xi_{jk}}{2} S X_j X_k
\ee
where the $G_\text{SM}$ singlet $X$ has a Majorana fermionic component.

The discrete symmetry of the $U(1)'$ charged hidden sector is
\be
Z_2^{hid}:~~ g_2^{hid} = U_2^u .
\ee
We define $\cal U$ as $X$ number similar to ${\cal B}$ and ${\cal L}$.
The discrete charge of $U_2^u$ is $q = -u {\cal U} \md 2$.
Since the MSSM fields are neutral under $Z_2^{hid}$, assuming that all possible exotic fields are heavier than the lightest $X$, the lightest $U$-parity ($U_2$) particle can be only the hidden sector field $X$.
The LUP is stable by the $U$-parity, and it can be either fermionic or scalar component (whichever lighter) of the $X$.

A discrete symmetry $Z_N$ with $N = N_1 N_2$ is isomorphic to the product of two discrete symmetries $Z_{N_1}$ and $Z_{N_2}$
\be
Z_{N} = Z_{N_1} \times Z_{N_2} \quad
\ee
if $N_1$ and $N_2$ are coprime (i.e. their greatest common divisor is $1$).
Then both of the stable particles under each discrete symmetry $Z_{N_1}$ and $Z_{N_2}$ are still stable under $Z_N$.

\begin{figure}[tb]
\includegraphics[width=0.48\textwidth]{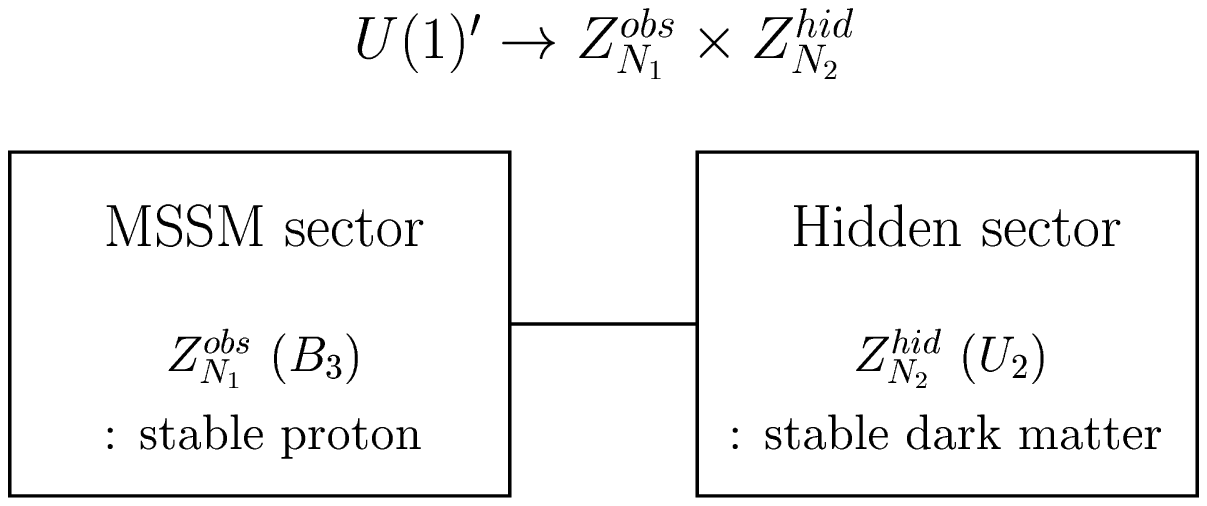}
\caption{A common $U(1)'$ gauge symmetry provides discrete symmetries to both sectors stabilizing proton and dark matter.
A unified picture of a single $U(1)'$ gauge symmetry over observable sector ($G_\text{SM}$ charged fields) and hidden sector ($G_\text{SM}$ singlet fields) that addresses the $\mu$-problem, proton stability, and dark matter stability arises.
}
\label{fig:diagram}
\end{figure}

Consider a product of $Z_{N_1}^{obs}$ and $Z_2^{hid}$
\bea
Z_{2 N_1}^{tot} :~~ g^{tot}_{2N_1} &=& g_{N_1}^{obs} \times g_2^{hid} \\
                                   &=& B_{N_1}^b L_{N_1}^\ell \times U_2^u \\
                                   &=& B_{2N_1}^{2b} L_{2N_1}^{2\ell} U_{2N_1}^{N_1u}
\eea
with $N_1$ coprime to $2$.
Since both sectors are charged under the $U(1)'$ gauge symmetry, a single $U(1)'$ which has $Z_{2N_1}^{tot}$ as its residual discrete symmetry provides discrete symmetries to both sectors.
Therefore, the stable lightest baryonic matter (proton) and dark matter (LUP) can be guaranteed by the common $U(1)'$ symmetry without $R$-parity (see Figure \ref{fig:diagram}).
This $U(1)'$ is further motivated to solve the $\mu$-problem as we discuss later.

%%%%%%%%%%%%%%%%%%%%%%%%%%%% FIGURE %%%%%%%%%%%%%%%%%%%%%%%%%%%%%%%%
%%%%%%%%%%%%%%%%%%%%%%%%%%%%%%%%%%%%%%%%%%%%%%%%%%%%%%%%%%%%%%%%%%%

%--------------------------------------------------------------------------------
\section{Example}
%--------------------------------------------------------------------------------
Here, we will consider the simplest example of the $R$-parity free model with LUP dark matter.
We assume the minimal discrete symmetries that guarantee the absolute stabilities to the proton and dark matter, i.e. $B_3$ for the MSSM sector discrete symmetry and $U_2$ for the hidden sector discrete symmetry.
The remnant discrete symmetry of the $U(1)'$ is therefore
\be
Z_6^{tot}:~~ g_6^{tot} = B_6^2 U_6^3
\ee
and its total discrete charge is given by $q = 2 q_B + 3 q_U \md 6$.
\be
\larray{
q[Q]=0 & q[U^c]=-2 & q[D^c]=2 \\
q[L]=-2 & q[E^c]=-2 & q[N^c]=0 \\
q[H_u]=2 & q[H_d]=-2 & q[X]=-3
}
\ee

Now, the proton and dark matter are absolutely stable.
However, to be a viable dark matter candidate, the LUP should satisfy the relic density and the direct detection constraints simultaneously.
In Ref. \cite{Hur:2007ur}, it was shown that when the LUP is effectively the only dark matter (i.e. $\Omega_\text{LUP} \gg \Omega_\text{LSP}$), it can still satisfy both relic density and direct detection constraints for a wide range of parameter space.

Because of the $R$-parity violation there are interesting predictions such as sneutrino resonance at collider experiments.
A slowly decaying TeV scale LSP may be still long-lived enough to be a dark matter candidate if the $R$-parity violating coupling is sufficiently small.
The requirement that the LSP lifetime be longer than the Universe age imposes extremely severe constraint on the $R$-parity violating coefficient, and any observation of the $R$-parity violating signals at collider would likely rule out the LSP as a viable dark matter candidate.
With the LUP, however, such a $R$-parity violating signal is not constrained by the dark matter data, since the $U$-parity that stabilizes the LUP dark matter does not constrain the ${\cal L}$ or ${\cal B}$ violating processes.
Also many superparticles such as charged sleptons and left-handed sneutrinos that were disfavored as the lightest one by the dark matter constraint are now also allowed to be the LSP as they can decay through $R$-parity violating processes.

%--------------------------------------------------------------------------------
\section{Beyond discrete symmetries}
\label{sec:beyond}
%--------------------------------------------------------------------------------
Some issues need discussions with the $U(1)'$ gauge symmetry since its discrete symmetry cannot help.
A TeV scale $U(1)'$ gauge symmetry can solve the $\mu$-problem by replacing $\mu H_u H_d$ with $h_s S H_u H_d$ through the $U(1)'$ charge assignment
\be
z[H_u]+z[H_d] \ne 0, \quad z[S]+z[H_u]+z[H_d] = 0 .
\ee
When $S$ gets a vev, an effective $\mu$ parameter
\be
\mu_\text{eff} = h_s \left< S \right>
\ee
at TeV scale is dynamically generated resolving the $\mu$-problem \cite{UMSSM}.
This does not change any discrete symmetry argument of this letter since $q[S] = 0$.

This mechanism requires some exotic colors due to the $[SU(3)_C]^2-U(1)'$ anomaly cancellation (see Ref. \cite{Lee:2007fw} and references therein).
We assume any exotic fields are heavier than the proton and the lightest $X$ field so that they are not stable due to the discrete symmetry.
The colored exotics may ruin the gauge coupling unification \cite{Morrissey:2005uz}, which we do not address in our rather phenomenological approach.

With the $U(1)'$ charge assignment, tightly constrained values of the $R$-parity violating coefficients can be also explained.
For example, replacing $\lambda L L E^c$ with a nonrenormalizable term $(\frac{S}{M})^n L L E^c$ can provide a naturally suppressed effective coefficient
\be
\lambda_\text{eff} = \left(\frac{\left<S\right>}{M}\right)^n .
\ee

An explicit construction of the $U(1)'$ model is beyond the scope of this letter, but a general method of finding the $U(1)'$ charges in the $R$-parity violating models for the MSSM fields was investigated in Refs. \cite{Lee:2007fw,Lee:2007qx} although the exotic fields part is highly model-dependent.

%--------------------------------------------------------------------------------
\section{Possible cure of gravitino problem}
%--------------------------------------------------------------------------------
LUP can be a natural and appealing solution to the gravitino problem in a typical\footnote{With an additional $U(1)'$ gauge symmetry, a novel mechanism such as new gaugino messenger scenario is possible which can ameliorate this problem substantially \cite{Langacker:2007ac}.} gauge mediated SUSY breaking (GMSB) scenario \cite{Giudice:1998bp}.
In the gravity mediated SUSY breaking scenario, there are multiple candidates to be the LSP before any experimental constraint is applied.
In the GMSB, gravitino is almost inevitably the only LSP candidate due to the hierarchy between the messenger scale and Planck scale.
The gravitino contribution to the critical density \cite{Pagels:1981ke} is approximately given by
\be
\Omega_{3/2} h^2 \sim \frac{m_{3/2}}{1~\text{keV}} .
\ee
If the light gravitino is the stable dark matter, its mass is constrained to be $m_{3/2} \sim {\cal O}(\text{keV})$.
At the structure formation it would have been a warm dark matter, which cannot explain the matter power spectrum \cite{Viel:2005qj}.
When the LUP is a dominant dark matter with lighter gravitino LSP ($m_{3/2} \ll 1~\text{keV}$) as a subdominant or negligible dark matter (due to the smallness of the coupling and mass, it may be still long-lived in the absence of $R$-parity), the matter power spectrum can be explained without adopting non-standard cosmology.
The next-to-lightest superparticle (NLSP) will decay into the SM particles through the $R$-parity violating processes.
With LUP dark matter, gravitino LSP can be also heavier than ${\cal O}(\text{keV})$,
decaying through the $R$-parity violating processes and contributing negligibly to the relic density.
More detailed analysis should be performed to fully understand the constraints on the gravitino LSP.

%--------------------------------------------------------------------------------
\section{Summary and Conclusions}
%--------------------------------------------------------------------------------
%==================== TABLE ====================================================================
\begin{table}[tb]
\begin{center}
\begin{tabular}{|l||l|l|}
\hline
               & $R$-parity                & $U(1)' \to B_3 \times U_2$ \\
\hline \hline
$\mu$-problem  & not addressed             & solvable \\
\hline
proton         & unstable with dim 5 op.   & stable \\
\hline
dark matter    & stable (LSP)              & stable (LUP) \\
\hline
gravity effect & violation (unless gauged) & no violation \\
\hline
RPV signals    & impossible                & possible \\
\hline
light $\widetilde G$ problem & not addressed & solvable \\
\hline
\end{tabular}
\end{center}
\caption{Comparison of supersymmetric models with the $R$-parity and the $U(1)'$ as a SUSY companion symmetry. \label{tab:comparison}}
\end{table}
%=============================================================================================

Though $R$-parity has been a widely accepted stability mechanism for the proton and the LSP dark matter candidate, it has shortcomings to be a fulfilling TeV scale SUSY companion symmetry.
We proposed the TeV scale $U(1)'$ gauge symmetry as an alternative to the $R$-parity.
With residual discrete symmetries for both the MSSM sector and the hidden sector, the $U(1)'$ alone can guarantee the absolute or sufficient stabilities of the proton and the hidden sector dark matter candidate (LUP).
Thus, the LUP can be an attractive dark matter candidate in the $R$-parity violating models.
The usual gravitino problem of the GMSB models may be also avoided with the LUP dark matter in the absence of the $R$-parity.

Phenomenological implications of the LUP dark matter scenario are distinguishable from the LSP dark matter scenario including direct detection and collider signals, and it will be worth to investigate them in detail.
Table \ref{tab:comparison} summarizes some of the differences between the models with the $R$-parity and the $U(1)'$.
Since the $R$-parity (or the matter parity) should be also gauged to the $U(1)_{B-L}$ to be protected from the Plank scale physics, using the $U(1)'$ is not introducing {\em more} new physics, but replacing one $U(1)$ with another one.

Though we assumed only Majorana type hidden sector fields, an extension to the Dirac type is straightforward, and more general $Z_N^{hid}$ (with $N\ge2$) would be possible.
This issue as well as construction of the anomaly-free $U(1)'$ models will be studied in subsequent publications.

%%%%%%%%%%%%%%%%%%%%%%%%%%%%%
\begin{acknowledgments}
%%%%%%%%%%%%%%%%%%%%%%%%%%%%%
\vspace{5pt}
This work is supported by the Department of Energy under grant DE-FG02-97ER41029.
I am grateful to T. Hur, C. Luhn, K. Matchev and S. Nasri for manuscript reading and useful discussions. 
\end{acknowledgments}

%%%%%%%%%%%%%%%%%%%%%%%%%%%%%

%=======================================================================
% END THE DOCUMENT
%=======================================================================

\end{document}